\documentclass[11pt,prd,aps,floats,preprintnumbers]{revtex4}
\usepackage{amsmath,amssymb}
\usepackage{graphicx}
\usepackage{graphics}
\usepackage{epsfig}
\usepackage{latexsym,oldgerm}
\def\beq{\begin{equation}}
\def\eeq{\end{equation}}
\def\bea{\begin{eqnarray}}
\def\eea{\end{eqnarray}}

\begin{document}
\small \preprint{SU-4252-895 \vspace{1cm}}
\setlength{\unitlength}{1mm}
\title{Spin $j$ Dirac Operators on the Fuzzy $2$-Sphere}
\author{A. P. Balachandran$^{a,b}$}\thanks{bal@phy.syr.edu}\thanks{C\'{a}tedra de
Excelencia}
\author{Pramod Padmanabhan$^{a}$}\thanks{ppadmana@syr.edu}
\affiliation{$^{a}$Department of Physics, Syracuse University,
Syracuse, NY 13244-1130, USA} \affiliation{$^{b}$Departamento de
Matem´aticas, Universidad Carlos III de Madrid, 28911 Legan\'{e}s,
Madrid, Spain}

\begin{abstract}
\begin{center}
{\small ABSTRACT}
\end{center}
The spin $\frac{1}{2}$ Dirac operator and its chirality operator on
the fuzzy $2$-sphere $S^2_F$ can be constructed using the
Ginsparg-Wilson(GW) algebra ~\cite{Fuzzy}. This construction
actually exists for any spin $j$ on $S^2_F$, and have continuum
analogues as well on the commutative sphere $S^2$ or on
$\mathbb{R}^{2}$. This is a remarkable fact and has no known
analogue in higher dimensional Minkowski spaces. We study such
operators on $S^2_F$ and the commutative $S^2$ and formulate
criteria for the existence of the limit from the former to the
latter. This singles out certain fuzzy versions of these operators
as the preferred Dirac operators. We then study the spin $1$ Dirac
operator of this preferred type and its chirality on the fuzzy
$2$-sphere and formulate its instanton sectors and their index
theory. The method to generalize this analysis to any spin $j$ is
also studied in detail.
\end{abstract}
\maketitle

\section{Introduction}

The Dirac and chirality operators are central for fundamental
physics and also in noncommutative geometry, where it is used to
formulate metrical, differential geometric and bundle-theoretic
ideas following Connes' approach~\cite{AC}.

 The theory of these operators on the fuzzy sphere $S_F^2$ can be
formulated using the Ginsparg-Wilson(GW) algebra, or the approach of
~\cite{KP,Do} The GW algebra was originally encountered in the
context of lattice gauge theories ~\cite{GW} where it was formulated
in order to avoid the fermion doubling problem. The fact that this
algebra appears naturally in the fuzzy case is interesting. In
particular we shall see that it provides a way to formulate the
Dirac and chirality operators for any non-zero spin. The latter in
turn leads to a Dirac-like equation for any spin on $S^{2}$ and
$\mathbb{R}^{2}$ with its associated chirality operator.

 We shall hereafter refer to these Dirac-like equations and
their chiralities just as Dirac and chirality operators.

 These Dirac and chirality operators remind one of the
Duffin-Kemmer, Rarita-Schwinger and Bargmann-Wigner equations. The
relation between these well-known equations and those found in this
paper remain to be explored.

 In section $2$, we establish our notation for $S_F^2$ and recall the
earlier formulation of the GW algebra and the fuzzy Dirac and
chirality operators for spin $\frac{1}{2}$. In section $3$, we
examine the ambiguities in the construction of the fuzzy spin
$\frac{1}{2}$ Dirac operator and study their continuum limits as
well. When these limits exist, the resultant continuum operators are
unitarily equivalent.

 Section $4$ gives a procedure to construct these operators in the
continuum. These will act as a guide in taking the limit of their
corresponding fuzzy versions, thereby fixing the fuzzy Dirac and
chirality operators. With this in mind, we explicitly construct the
spin $1$ Dirac and chirality operators in the continuum.

 Then in section $5$ we go on to construct the fuzzy versions of the
spin $1$ Dirac operator by the construction of their GW algebras in
such a manner that the continuum limits exist.

 In section $6$, guided by the analysis in section $5$, the GW systems
and hence their Dirac and chirality operators are constructed for
any spin in such a manner that their continuum limits exist. Crucial
observations of terms arising in such computations are made and the
procedure for taking their continuum limit is discussed in detail.

 In section $7$ we summarize our rules for finding the fuzzy Dirac
and chirality operators. We also prove a claim which unambiguously
fixes the fuzzy Dirac and chirality operators for all spins.

 Instanton sectors can be formulated in the algebraic language in
terms of projective modules ~\cite{ApbYd}. There is a natural
adaptation of this idea to $S_F^2$ for scalar and spin $\frac{1}{2}$
fields ~\cite{Fuzzy,ApG}. The index theory has also been established
in the latter case. In section 8, we generalize this construction to
any spin and their Dirac and chirality operators on $S_F^2$ and also
establish their index theory.

 In section $9$ we present our conclusions.

\section{The fuzzy sphere and its GW algebra}
 The algebra for the fuzzy sphere is
characterized by a cut-off angular momentum $L$ and is the full
matrix algebra $Mat(2L+1)\equiv M_{2L+1}$ of $(2L+1)\times (2L+1)$
matrices. They can be generated by the $(2L+1)$-dimensional
irreducible representation (IRR) of $SU(2)$ with the standard
angular momentum basis. The latter is represented by the angular
momenta $L^L_i$ acting on the left on $Mat(2L+1)$: If $\alpha\in
Mat(2L+1)$, \beq{L_{i}^{L}\alpha=L_{i}\alpha}\eeq
\beq{[L_{i}^{L},L_{j}^{L}]=i\epsilon_{ijk}L_{k}^{L}}\eeq
\beq{(L_{i}^{L})^2=L(L+1)\mathbf{1}}\eeq where $L_i$ are the
standard angular momentum matrices for angular momentum $L$.

 We can also define right angular momenta $L_i^R$:
\beq{L_{i}^{R}\alpha=\alpha L_{i}, \alpha\in M_{2L+1}}\eeq
\beq\label{commu}{[L_{i}^{R},L_{j}^{R}]=-i\epsilon_{ijk}L_{k}^R}\eeq
\beq{(L_{i}^{R})^2=L(L+1)\mathbf{1}}\eeq We also have
\beq{[L_i^L,L_j^R]=0.}\eeq

 The operator $\mathcal{L}_i=L_i^L-L_i^R$ is the fuzzy version of orbital
angular momentum. They satisfy the $SU(2)$ angular momentum algebra
\beq{[\mathcal{L}_i,\mathcal{L}_j]=i\epsilon_{ijk}\mathcal{L}_k}\eeq

 In the continuum, $S^2$ can be described by the unit vector
$\hat{x}\in S^2$, where $\hat{x}.\hat{x}=1$. Its analogue on $S_F^2$
is $\frac{L_i^L}{L}$ or $\frac{L_I^R}{L}$ such that
\beq{\lim_{L\rightarrow\infty}\frac{L_i^{L,R}}{L}=\hat{x}_i.}\eeq
This shows that $L_i^{L,R}$ do not have continuum limits. But
$\mathcal{L}_i=L_i^L-L_i^R$ does and becomes the orbital angular
momentum as $L\rightarrow\infty$:
\beq{\lim_{L\rightarrow\infty}L_i^L-L_i^R=-i(\overrightarrow{r}\wedge\overrightarrow{\nabla})_i
.}\eeq

\subsection*{The GW Algebra}
 In algebraic terms, the GW algebra $\mathcal{A}$ is the unital $\ast$
algebra over $\mathbf{C}$ ,generated by two $\ast$-invariant
involutions $\Gamma, \Gamma'$.
 \beq \label{GW}\mathcal{A}=\{\Gamma,\Gamma'\ :\Gamma^2=\Gamma'^2=1\
,\Gamma^*=\Gamma\ ,\Gamma'^*=\Gamma'\}\eeq

 In any $\ast$ -representation on a Hilbert space,
$\ast$ becomes the adjoint $\dag$.

 Each representation of Eq.(\ref{GW}) is a particular realization of the GW algebra. Representations
of interest in fuzzy physics are generally reducible.

\subsection*{The Dirac Operator from GW Algebra}

Consider the following two elements constructed out of $\Gamma,
\Gamma'$: \beq {\Gamma_1=\frac{1}{2}(\Gamma+\Gamma'),}\eeq
\beq{\Gamma_2=\frac{1}{2}(\Gamma-\Gamma').}\eeq It follows from
Eq.(\ref{GW}) that $\{\Gamma_1,\Gamma_2\}=0$. This suggests that for
suitable choices of $\Gamma$, $\Gamma '$, one of these operators may
serve as the Dirac operator and the other as the chirality operator
provided they have the right continuum limits after suitable
scaling. This is indeed the case as we now show for the fuzzy spin
$\frac{1}{2}$ Dirac and chirality operators.

\subsection*{The Fuzzy Dirac Operator: Spin $\frac{1}{2}$}
 The construction is based on the GW algebra of~\cite{ApGi,ApbTrg}. First we
note that if $P$ is a projector, then, \beq{P^2=P}\eeq and
$\gamma=2P-1$ is an idempotent: \beq{\gamma^2=1.}\eeq

 We now construct $\Gamma$, $\Gamma'$ from suitable projectors.

 Consider $Mat(2L+1)\otimes\mathbb{C}^2$. The spin $\frac
{1}{2}$ IRR of $SU(2)$ acts on $\mathbb{C}^2$. It has the standard
Lie algebra basis $\frac{\sigma_i}{2}$, $\sigma_i$ being the Pauli
matrices. The projector coupling the left angular momentum and this
spin $\frac{1}{2}$ to its maximum value $L+\frac{1}{2}$ is
\beq{P^L_{L+\frac{1}{2}}=\frac{\vec{\sigma}.\vec{L}^L+L+1}{2L+1}.}\eeq
Hence the corresponding idempotent is
\beq{\Gamma_{L+\frac{1}{2}}^L=\frac{\vec{\sigma}.\vec{L}^L+\frac{1}{2}}{L+\frac{1}{2}}.}\eeq
The projector $P^R_{L+\frac{1}{2}}$ coupling the right angular
momentum and spin $\frac{1}{2}$ to $L+\frac{1}{2}$ is obtained by
changing $\vec{L}^L$ to $-\vec{L}^R$ in the above expression:
\beq{P^R_{L+\frac{1}{2}}=\frac{-\vec{\sigma}.\vec{L}^R+L+1}{2L+1}.}\eeq
The minus sign is because of the minus sign in Eq.(\ref{commu}).

The corresponding idempotent is
\beq{\Gamma^R_{L+\frac{1}{2}}=\frac{-\vec{\sigma}.\vec{L}^R+\frac{1}{2}}{L+\frac{1}{2}}.}\eeq

Identifying $\Gamma_{L+\frac{1}{2}}^{L,R}$ with $\Gamma$, $\Gamma'$,
we get
\beq{\Gamma_1=\frac{1}{2}\left[\frac{\vec{\sigma}.\vec{\mathcal{L}}+1}{L+\frac{1}{2}}\right]}\eeq
and
\beq{\Gamma_2=\frac{1}{2}\left[\frac{\vec{\sigma}.(\vec{L}^L+\vec{L}^R)}{L+\frac{1}{2}}\right]}\eeq

Now as $L\rightarrow\infty$,\beq \label{D1/2}{
2L\Gamma_1\rightarrow\vec{\sigma}.\vec{\mathcal{L}}+1}\eeq and
\beq{\Gamma_2\rightarrow\vec{\sigma}.\hat{x}.}\eeq These are the
correct Dirac and chirality operators on $S^2$ and so we can regard
$2L\Gamma_1$ as the fuzzy Dirac operator (upto a finite scaling) and
$\Gamma_2$ as its chirality operator.

\section{Ambiguities in the fuzzy spin $\frac{1}{2}$ Dirac and Chirality Operators}
 Having looked at the construction of the spin $\frac{1}{2}$ Dirac
operator as given in ~\cite{Fuzzy}, we now consider other
possibilities for constructing the same Dirac operator. This
observation turns out to be crucial in finding the Dirac operator
for higher spins.

 The projectors $P^{L,R}_{L+\frac{1}{2}}$ are not the only
projectors with rotational invariance. We can also consider the two
projectors to the $L-\frac{1}{2}$ space, obtained by coupling the
left and right angular momenta $L_i^{L,R}$ and spin $\frac{1}{2}$.
These are,
\beq{P_{L-\frac{1}{2}}^L=-(\frac{\vec{\sigma}.\vec{L}^L-L}{2L+1}),}\eeq
and
\beq{P_{L-\frac{1}{2}}^R=-(\frac{-\vec{\sigma}.\vec{L}^R-L}{2L+1}).}\eeq
This gives us two new generators, $\Gamma_{L-\frac{1}{2}}^{L,R}$, to
the GW algebra. Thus there are a total of four rotationally
invariant idempotents which we list in the following table \beq
\label{P1/2}{P^{L,R}_{L+\frac{1}{2}}: \ \ \ \
\Gamma^{L}_{L+\frac{1}{2}} \ \ \ \ \Gamma^{R}_{L+\frac{1}{2}}}\eeq
\beq\label{P-1/2}{P^{L,R}_{L-\frac{1}{2}}: \ \ \ \
\Gamma^{L}_{L-\frac{1}{2}} \ \ \ \ \Gamma^{R}_{L-\frac{1}{2}}}\eeq
The negatives of these idempotents are also idempotents, but that is
a trivial ambiguity.

 Now a GW algebra is generated by any pair from this table. However
if we adopt the two left or the two right as $\Gamma$ and $\Gamma'$,
then $\Gamma_1$ and $\Gamma_2$ have no suitable continuum limit. We
can see this from choosing as our generators either
$\Gamma^L_{L\pm\frac{1}{2}}$ or $\Gamma^R_{L\pm\frac{1}{2}}$. We
observe that $\Gamma^L_{L+\frac{1}{2}}=-\Gamma^L_{L-\frac{1}{2}}$
and $\Gamma^R_{L+\frac{1}{2}}=-\Gamma^R_{L-\frac{1}{2}}$, which as
remarked above is a trivial ambiguity. So clearly we cannot
construct suitable GW algebras from such pairs of idempotents.

 But if we now use the two operators $\Gamma^{L}_{L+\frac{1}{2}}$ and
$\Gamma^{R}_{L-\frac{1}{2}}$ and consider the combination
$(L+\frac{1}{2})(\Gamma^{L}_{L+\frac{1}{2}}-\Gamma^{R}_{L-\frac{1}{2}})$,
we get the Dirac operator given in Eq.(\ref{D1/2}).  As we saw
earlier in section $2$ ~\cite{Fuzzy}, this Dirac operator is found
by adding $\Gamma^{L}_{L+\frac{1}{2}}$ and
$\Gamma^{R}_{L+\frac{1}{2}}$ and scaling as $L\rightarrow\infty$.
The corresponding chirality operator is got from
$\frac{\Gamma^{L}_{L+\frac{1}{2}}+\Gamma^{R}_{L-\frac{1}{2}}}{2}$ as
this goes to the correct limit as $L\rightarrow\infty$ which is
$\sigma.\hat{x}$. The other possibility of combining
$\Gamma^{L}_{L-\frac{1}{2}}$ and $\Gamma^{R}_{L+\frac{1}{2}}$ also
exists and it is easy to see that
$-(L+\frac{1}{2})(\Gamma^{L}_{L-\frac{1}{2}}-\Gamma^{R}_{L+\frac{1}{2}})$
also goes to the Dirac operator given by Eq.(\ref{D1/2}) while
$\frac{\Gamma^{L}_{L-\frac{1}{2}}+\Gamma^{R}_{L+\frac{1}{2}}}{2}$
goes to the corresponding chirality operator. This exhausts all the
possible combinations.

 We again note here that we can only construct our desired Dirac and chirality operators by
choosing one $\Gamma$ from the second column and one from the third
column of Eq.(\ref{P1/2}) and Eq.(\ref{P-1/2}) as we will not get a
differential operator in the continuum if we choose them from the
same column.

 The fact that there exist all these possibilities for combining various generators
of the GW algebra for obtaining the fuzzy Dirac and chirality
operators imply that we should take care while writing the
corresponding versions of higher spin Dirac and chirality operators
as not all of them may go to correct continuum limits. In the case
of spin $\frac{1}{2}$, all the possibilities go to the correct
continuum limit, but as we shall soon see, this fails in the case of
higher spins. This calls for a rule to construct the fuzzy versions
of these operators, which we shall formulate after studying the spin
$1$ case in detail. We shall also see later that this becomes
essential for finding the Dirac operators in the continuum for
higher spins.

 But there {\textit{are}} more substantial ambiguities to consider.
There are other operators in the GW algebra which can serve as Dirac
and chirality operators~\cite{ApbTrg,Fuzzy}. For example there are
those which give the Dirac and chirality operators of the Watamuras
~\cite{WW} on $S^F_2$. As shown in ~\cite{ApbTrg}, in the continuum
limit, the corresponding operators are unitarily equivalent to
Eq.(\ref{D1/2}). We will not pursue such ambiguities further here.

 In the next section we will see how to construct the Dirac operator
and chirality operator on $S^2$ for spin $\frac{1}{2}$ and spin $1$.

\section{The Dirac and Chirality Operators on $S^{2}$}
We can construct a set of anti-commuting operators and call them the
Dirac and chirality operators after checking that they have the
right properties. Consider
\beq\label{GD}{D=(\Sigma_{i}-\gamma\Sigma_{i}\gamma)(\mathcal{L}_{i}+\Sigma_{i})}\eeq
where $\gamma$ satisfies $\gamma^{2}=1$ and $\gamma\dag=\gamma$.
$\vec{\Sigma}$ is the spin $j$ representation of $SU(2)$. It is easy
to check that this form of $D$ in Eq.(\ref{GD}) implies that
\beq{\{D,\gamma\}=0}\eeq as $\gamma$ commutes with the total angular
momentum $J_{i}=\mathcal{L}_{i}+\Sigma_{i}$. This follows from the
following operator identity: \beq{\{A,BC\}=\{A,B\}C-B[A,C]}\eeq Thus
$D$ and $\gamma$ are Dirac and chirality operators.

\subsection*{D and $\gamma$ for  the Spin $\frac{1}{2}$ case}
Let us now explicitly construct $D$ and $\gamma$ for the spin
$\frac{1}{2}$ case.

In the fuzzy case $\vec{\sigma}.\vec{L}^L=L$ on the $L+\frac{1}{2}$
space and $\vec{\sigma}.\vec{L}^L=-(L+1)$ on the $L-\frac{1}{2}$
space. Thus taking their continuum limits gives us
$\vec{\sigma}.\hat{x}=\pm1$ on these two spaces. An alternative way
to find the eigenvalues of $\vec{\sigma}.\hat{x}$ without taking
continuum limits of the fuzzy case is by noting that we can choose
the direction of $\hat{x}$ to be along the third direction, which
implies the eigenvalues of $\vec{\sigma}.\hat{x}$ are just the
eigenvalues of $\sigma_3$ namely $\pm1$. This will be used
extensively when we generalize to higher spins.

Using $\vec{\sigma}.\hat{x}$,  we can construct the projectors onto
the two spaces with $\vec{\sigma}.\hat{x}=\pm1$:
\beq\label{P1C}{P_{1}=\frac{1+\vec{\sigma}.\hat{x}}{2}}\eeq and
\beq\label{P-1C}{P_{-1}=\frac{1-\vec{\sigma}.\hat{x}}{2}}\eeq Now
for any projector $P$, $1-2P$ is an idempotent:
\beq{(1-2P)^2=1.}\eeq Thus from Eq.(\ref{P1C}) and Eq.(\ref{P-1C}),
we can read off the two chirality operators as
$\pm\vec{\sigma}.\hat{x}$.

The Dirac operators corresponding to these two chirality operators
are the same due to the form of the Dirac operator given by
Eq.(\ref{GD}).

We can compute $D$ using the algebra of the Pauli matrices. That
gives us
\beq{\sigma_{i}-(\vec{\sigma}.\hat{x})\sigma_{i}(\vec{\sigma}.\hat{x})=\sigma_{i}-x_{i}(\vec{\sigma}.\hat{x})}\eeq
and thus from Eq.(\ref{GD}),
\beq{D=\vec{\sigma}.\vec{\mathcal{L}}+\frac{1}{2}}\eeq which is the
well-known continuum Dirac operator for spin $\frac{1}{2}$ on
$S^2$~\cite{Jaya}

\subsection*{D and $\gamma$ on $S^2$ for the spin $1$ case}
 In a similar fashion we can find the chirality operators in the
continuum for the spin $1$ case by noting that the eigenvalues of
$\vec{\Sigma}.\hat{x}$ are $\pm1$ and $0$. We then write the
projectors to the spaces where $\vec{\Sigma}.\hat{x}$ takes these
three values and by writing these projectors as $\frac{1+\gamma}{2}$
we can read off the three chirality operators. They are
\beq\label{Ch0}{\gamma_{1}=1-2(\vec{\Sigma}.\hat{x})^{2},}\eeq
\beq\label{Ch1}{\gamma_{2}=(\vec{\Sigma}.\hat{x})^{2}+(\vec{\Sigma}.\hat{x})-1,}\eeq
\beq\label{Ch-1}{\gamma_{3}=(\vec{\Sigma}.\hat{x})^{2}-(\vec{\Sigma}.\hat{x})-1.}\eeq

The Dirac operator corresponding to Eq.(\ref{Ch0}) is found to be
\beq\label{S1D}{D_{1}=\vec{\Sigma}.\vec{\mathcal{L}}-(\vec{\Sigma}.\hat{x})^{2}+2.}\eeq
The ones corresponding to the other chirality operators are
unitarily equivalent to this one. The corresponding unitary operator
transforms the eigenspace of $\vec{\Sigma}\hat{x}$ with eigenvalue
$0$ to either of the other eigenvalues. It is easy to write down the
unitary operator connecting these chiralities if we take $\hat{x}$
to be in the third direction. If this is the case the three
chiralities become \beq\label{c1m}{\gamma_1=\left(\begin{array}{ccc} -1 & 0 & 0\\
0 & 1 & 0\\ 0 & 0 & -1\end{array}\right),}\eeq \beq\label{c2m}{\gamma_2=\left(\begin{array}{ccc} 1 & 0 & 0\\
0 & -1 & 0\\ 0 & 0 & -1\end{array}\right)}\eeq and \beq\label{c3m}{\gamma_3=\left(\begin{array}{ccc} -1 & 0 & 0\\
0 & -1 & 0\\ 0 & 0 & 1\end{array}\right).}\eeq The unitary matrix
transforming Eq.(\ref{c1m}) to Eq.(\ref{c2m}) is
\beq{U=\left(\begin{array}{ccc} 0 & -i & 0\\ i & 0 & 0\\ 0 & 0 &
-1\end{array}\right).}\eeq

We do not know how the general unitary transform between the three
chiralities will be. We suspect it to be an operator of the form
$e^{iD}$ where $D$ is the Dirac operator.

Our fuzzy Dirac and chirality operators will have these as their
continuum limits.

The Dirac operator in Eq.(\ref{S1D}) is found using the algebra of
the spin $1$ matrices ~\cite{Va} which is used to simplify
\beq{[\Sigma_{i}-(1-2(\vec{\Sigma}.\hat{x})^{2})\Sigma_{i}(1-2(\vec{\Sigma}.\hat{x}))^{2}](\mathcal{L}_i+\Sigma_{i}).}\eeq
We simplify the term in the square bracket after writing it in the
form
\beq\label{18}{[2\Sigma_{i}(\vec{\Sigma}.\hat{x})^{2}+(\vec{\Sigma}.\hat{x})^{2}\Sigma_{i}-4(\vec{\Sigma}.\hat{x})^{2}\Sigma_{i}(\vec{\Sigma}.\hat{x})^{2}].}\eeq
The first two terms in the above expression can be simplified using
\beq\label{19}{\Sigma_{i}\Sigma_{k}\Sigma_{j}=\frac{i}{3}\varepsilon_{ikj}+\frac{1}{2}(\delta_{ik}\Sigma_{j}+\delta_{kj}\Sigma_{i})+i\varepsilon_{ijm}Q_{km}}\eeq
where $Q_{km}$ is a symmetric tensor. This identity gives the sum of
the first two terms as
\beq{A+B=2\Sigma_{i}+2(\vec{\Sigma}.\hat{x})x_{i}}\eeq where $A$ and
$B$ are the first two terms in Eq.(\ref{18}). The identity in
Eq.(\ref{19}) can also be used to simplify the third term in
Eq.(\ref{18}) and we get
\beq{C=2\Sigma_{i}(\vec{\Sigma}.\hat{x})^{2}+2x_{i}(\vec{\Sigma}.\hat{x})-4i\varepsilon_{ikm}Q_{jm}(\vec{\Sigma}.\hat{x})^{2}x_{k}x_{j}.}\eeq
Using Eq.(\ref{19}), we can simplify this further to
\beq\label{22}{C=3x_i(\vec{\Sigma}.\hat{x})+\Sigma_i+2i\varepsilon_{ijm}Q_{km}x_kx_j-4i\varepsilon_{ikm}Q_{jm}(\vec{\Sigma}.\hat{x})^2x_kx_j.}\eeq

To evaluate this, we need to simplify the last term in the
expression. That can be done using the following identities:
\beq{\Sigma_{l}\Sigma_{n}=\frac{2}{3}\delta_{ln}+\frac{i}{2}\varepsilon_{lno}\Sigma_{o}+Q_{ln}}\eeq
and
\beq\begin{split}{Q_{jm}Q_{ln}&=\frac{1}{6}(\delta_{jl}\delta_{mn}+\delta_{jn}\delta_{lm}-\frac{2}{3}\delta_{jm}\delta_{ln})\\&-\frac{1}{4}(\delta_{jl}Q_{mn}+\delta_{jn}Q_{lm}+\delta_{mn}Q_{jl}+\delta_{ml}Q_{jn}-\frac{4}{3}\delta_{jm}Q_{ln}-\frac{4}{3}\delta_{ln}Q_{jm})
\\&+\frac{i}{8}(\delta_{jl}\varepsilon_{mnp}\Sigma_{p}+\delta_{jn}\varepsilon_{mlp}\Sigma_{p}+\delta_{ml}\varepsilon_{jnp}\Sigma_{p}+\delta_{mn}\varepsilon_{jlp}\Sigma_{p}).}\end{split}\eeq
On using these two identities, the last term in Eq.(\ref{22})
becomes
\beq{2i\varepsilon_{ikm}Q_{lm}x_{k}x_{l}-x_{i}(\vec{\Sigma}.\hat{x})+\Sigma_{i}}\eeq
This can then be substituted in Eq.(\ref{22}) to get
\beq{C=4x_{i}(\vec{\Sigma}.\hat{x}).}\eeq With this, we obtain the
following simple expression for $A+B-C$:
\beq{A+B-C=\Sigma_{i}-x_{i}(\vec{\Sigma}.\hat{x})}\eeq

Multiplying this with $(\vec{\mathcal{L}_i}+\vec{\Sigma_{i}})$ gives
the Dirac operator in Eq.(\ref{S1D}).

Next we write down the Dirac operators corresponding to the other
chirality operators.

The Dirac operators corresponding to Eq.(\ref{Ch1}) and
Eq.(\ref{Ch-1}) are found to be
\beq\label{s1d1}{D_{2}=(\vec{\Sigma}.\vec{\mathcal{L}}-(\vec{\Sigma}.\hat{x})^{2}+2)+2(\vec{\Sigma}.\hat{x})+\{\vec{\Sigma}.\vec{\mathcal{L}},\vec{\Sigma}.\hat{x}\}}\eeq
and
\beq\label{s1d-1}{D_{3}=(\vec{\Sigma}.\vec{\mathcal{L}}-(\vec{\Sigma}.\hat{x})^{2}+2)-2(\vec{\Sigma}.\hat{x})-\{\vec{\Sigma}.\vec{\mathcal{L}},\vec{\Sigma}.\hat{x}\}.}\eeq

These are found using the algebra of spin $1$ matrices~\cite{Va} as
before.

These are the continuum limits which guide us in finding the fuzzy
spin $1$ Dirac operators. This will be explained in the next section
where we discuss in detail the construction of the fuzzy spin $1$
Dirac operator.

\section{The Fuzzy Spin $1$ Dirac Operator}
 Consider $Mat(2L+1)\bigotimes\mathbb{C}^{3}$, where $Mat(2L+1)$ is
 the carrier space of spin $L\otimes L$ representation of $SU(2)$ acting on left and right and
 $\mathbb{C}^{3}$ is the carrier space of the spin $1$ representation
 of $SU(2)$. When a spin $L$ couples with spin $1$, we have three
 possible spaces labeled by the values of the total angular
 momentum $L+1,L$ and $L-1$. So we have six
 projectors and as in Eq.(\ref{P1/2}) and Eq.(\ref{P-1/2}) we can construct the corresponding generators of the GW algebra. Thus we have a table similar to the one in Eq.(\ref{P1/2}) and
 Eq.(\ref{P-1/2}):
  \beq\label{P1}{P^{L,R}_{L+1}: \ \ \ \ \Gamma_{L+1}^{L} \ \ \ \ \Gamma_{L+1}^{R}}\eeq \beq\label{P0}{P^{L,R}_{L}: \ \ \ \
\Gamma_{L}^{L} \ \ \ \ \Gamma_{L}^{R}}\eeq
\beq\label{P-1}{P^{L,R}_{L-1}: \ \ \ \ \Gamma_{L-1}^{L} \ \ \ \
\Gamma_{L-1}^{R}}\eeq The notation used here is similar to the one
used in section 3.

The three projectors corresponding to the left angular momentum
coupling to spin $1$ are
\beq{P^L_{L+1}=\frac{(\vec{\Sigma}.\vec{L}^L+L+1)(\vec{\Sigma}.\vec{L}^L+1)}{(L+1)(2L+1)}}\eeq
\beq{P^L_{L}=-\frac{(\vec{\Sigma}.\vec{L}^L-L)(\vec{\Sigma}.\vec{L}^L+L+1)}{L(L+1)}}\eeq
\beq{P^L_{L-1}=\frac{(\vec{\Sigma}.\vec{L}^L-L)(\vec{\Sigma}.\vec{L}^L+1)}{(2L+1)L}}\eeq
while the corresponding right projectors are obtained from above by
substituting $\vec{L}^L$ by $-\vec{L}^R$.

 Writing each projector as $\frac{1+\Gamma}{2}$ and $\vec{L}$ as $\vec{L}^L$ or
$-\vec{L}^R$, we can find the generators of the GW algebra for each
of the projectors above. Let us write down the relevant generators
whose combinations give the fuzzy Dirac and chirality operators
having the right continuum limits which we found in the previous
section.
\beq{\Gamma_{L+1}^{L}=\frac{2(\vec{\Sigma}.\vec{L}^L+L+1)(\vec{\Sigma}.\vec{L}^L+1)-(L+1)(2L+1)}{(L+1)(2L+1)}}\eeq
\beq\label{39}{\Gamma_{L+1}^{R}=\frac{2(-\vec{\Sigma}.\vec{L}^R+L+1)(-\vec{\Sigma}.\vec{L}^R+1)-(L+1)(2L+1)}{(L+1)(2L+1)}}\eeq
\beq\label{40}{\Gamma_{L-1}^{L}=\frac{2(\vec{\Sigma}.\vec{L}^L-L)(\vec{\Sigma}.\vec{L}^L+1)-L(2L+1)}{L(2L+1)}}\eeq
\beq{\Gamma_{L-1}^{R}=\frac{2(\vec{\Sigma}.\vec{L}^R+L)(\vec{\Sigma}.\vec{L}^R-1)-L(2L+1)}{L(2L+1)}}\eeq
We can immediately see that
$\frac{\Gamma_{L-1}^{L}\pm\Gamma_{L+1}^{R}}{2}$, are chirality and
Dirac operators (the latter upto an overall constant) for the fuzzy
sphere by checking their continuum limits. Thus as
$L\rightarrow\infty$,
\beq{\frac{\Gamma_{L-1}^{L}+\Gamma_{L+1}^{R}}{2}\rightarrow
(\vec{\Sigma}.\hat{x})^{2}-\vec{\Sigma}.\hat{x}-1,}\eeq which is a
chirality operator for the spin $1$ case in the continuum which we
encountered in the previous section. Also
\beq\label{lim}{\lim_{L\rightarrow\infty}
L(\frac{\Gamma_{L-1}^{L}-\Gamma_{L+1}^{R}}{2})=
-(\vec{\Sigma}.\vec{\mathcal{L}}-(\vec{\Sigma}.\hat{x})^{2}+2)}\eeq
is the corresponding Dirac operator as
$L(\frac{\Gamma_{L-1}^{L}-\Gamma_{L+1}^{R}}{2})$ anti-commutes with
$\frac{\Gamma_{L-1}^{L}+\Gamma_{L+1}^{R}}{2}$. The Dirac operator
got from the fuzzy case in Eq.(\ref{lim})is unitarily equivalent to
the one got in Eq.(\ref{s1d1}). This can be seen as a consequence of
the fact that the chiralities corresponding to these Dirac operators
are unitarily equivalent. Eq.(\ref{lim}) can be seen by substituting
the expressions for $\Gamma^L_{L-1}$ and $\Gamma^R_{L+1}$ from
Eq.(\ref{40}) and Eq.(\ref{39}) respectively and grouping terms
similar in the order of $\vec{L}^L$ and $\vec{L}^R$.

Here we note the order $L$ term in the expression
\beq{L\frac{((\vec{\Sigma}.\vec{L}^L)^{2}-(\vec{\Sigma}.\vec{L}^R)^{2})}{(L+1)(2L+1)}}\eeq
got by grouping the second order terms. As $L\rightarrow\infty$ this
term goes to $-\frac{(\vec{\Sigma}.\hat{x})}{2}$. This can be
understood easily by noting that
$[L^L_i,L^L_j]=i\varepsilon_{ijk}L^L_k$, produces first order terms
in $L$ and these commutators arise when we expand $L^L_iL^L_j$ as a
sum of a commutator and an anticommutator. However, this is just the
highest order term and this limit is not exact. We shall in fact see
later that the exact limit is different from this involving a first
order differential term thereby changing the form of the Dirac
operator. But we will show it to be unitarily equivalent to the
above Dirac operator.

Similarly we find the chirality and Dirac operators
$\frac{\Gamma_{L+1}^{L}+\Gamma_{L-1}^{R}}{2}$ for the fuzzy
sphere(the latter upto a constant) and their continuum limits.
\beq{\frac{\Gamma_{L+1}^{L}+\Gamma_{L-1}^{R}}{2}\rightarrow(\vec{\Sigma}.\hat{x})^{2}+\vec{\Sigma}.\hat{x}-1}\eeq
and
\beq{L(\frac{\Gamma_{L+1}^{L}-\Gamma_{L-1}^{R}}{2})\rightarrow(\vec{\Sigma}.\vec{\mathcal{L}}-(\vec{\Sigma}.\hat{x})^{2}+2)}\eeq
as $L\rightarrow\infty$. The Dirac operator got from the fuzzy case
in Eq.(\ref{lim})is unitarily equivalent to the one got in
Eq.(\ref{s1d-1}). Again this can be seen as a consequence of the
fact that the chiralities corresponding to these Dirac operators are
unitarily equivalent. We will remark more about this later.

 We can also see that $\gamma_{1}$ in Eq.(\ref{Ch0}) is
got by taking the continuum limit of
\beq{\frac{\Gamma_{L}^{L}+\Gamma_{L}^{R}}{2}}\eeq where
\beq{\Gamma^L_L=\frac{-2(\vec{\Sigma}.\vec{L}^L-L)(\vec{\Sigma}.\vec{L}^L+L+1)-L(L+1)}{L(L+1)}}\eeq
\beq{\Gamma^R_L=\frac{2(\vec{\Sigma}.\vec{L}^R+L)(-\vec{\Sigma}.\vec{L}^R+L+1)-L(L+1)}{L(L+1)}}\eeq
 This implies $L(\frac{\Gamma_{L}^{L}-\Gamma_{L}^{R}}{2})$
goes to the corresponding Dirac operator. Thus
$\frac{\Gamma_{L}^{L}+\Gamma_{L}^{R}}{2}$ and constant times
$\frac{\Gamma_{L}^{L}-\Gamma_{L}^{R}}{2}$ can also serve as
chirality and Dirac operators.

The continuum limit of the combination $\Gamma^R_L+\Gamma^L_{L+1}$
goes to $\vec{\Sigma}.\hat{x}-(\vec{\Sigma}.\hat{x})^2$ which is not
part of the chiralities we obtained in the continuum in section 4.
They are not unitarily to equivalent to any of those obtained in
section 4 either. Other combinations like
$\Gamma^R_L+\Gamma^L_{L-1}$ go to a chirality we do not have in the
continuum as formulated in section 4. The combinations anticommuting
with these namely $L(\Gamma^R_L-\Gamma^L_{L+1})$ and
$L(\Gamma^R_L+\Gamma^L_{L-1})$ do not have proper continuum limits,
in fact they diverge. Hence we discard these combinations.

\section{Generalizing to higher spins}
 The projectors to spaces, got by coupling $L$ to higher spins contain
more factors increasing the order in $\vec{L}^{L,R}$  and making the
expressions look complicated. We observe the kind of terms that can
emerge from simplifying these expressions and formulate rules to
take their continuum limits.

 We first carefully look at the spin $\frac{3}{2}$ case and use this to
generalize to terms emerging from higher spins. We have eight
projectors in this case which are $P^{L,R}_{L+\frac{3}{2}}$,
$P^{L,R}_{L+\frac{1}{2}}$, $P^{L,R}_{L-\frac{1}{2}}$,
$P^{L,R}_{L-\frac{3}{2}}$. We can construct the generators of the GW
algebra from each of these projectors and thus construct a table
similar to that shown in Eq.(\ref{P1})-Eq.(\ref{P-1}). From this
table, let us take the relevant $\Gamma$ operators whose combination
gives us the fuzzy Dirac operator. Consider
\beq{\Gamma^{L}_{L+\frac{3}{2}}=\frac{
(2\vec{\Sigma}.\vec{L}^L-L+3)(2\vec{\Sigma}.\vec{L}^L+L+4)(2\vec{\Sigma}.\vec{L}^L+3L+3)-6(L+1)(2L+3)(2L+1)}{6(L+1)(2L+1)(2L+3)}}\eeq
\beq{\Gamma^{R}_{L-\frac{3}{2}}=\frac{
(-2\vec{\Sigma}.\vec{L}^R-L+3)(-2\vec{\Sigma}.\vec{L}^R+L+4)(2\vec{\Sigma}.\vec{L}^R+3L)-6L(2L-1)(2L+1)}{6L(2L+1)(2L-1)}}\eeq
Now as $L\rightarrow\infty$,
\beq{\Gamma^{L}_{L+\frac{3}{2}}+\Gamma^{R}_{L-\frac{3}{2}}\rightarrow\frac{8(\vec{\Sigma}.\hat{x})^{3}-2\vec{\Sigma}.\hat{x}+12(\vec{\Sigma}.\hat{x})^{2}-27}{24}}\eeq
The Dirac operator corresponding to this can be got from taking the
continuum limit of
$L(\Gamma^{L}_{L+\frac{3}{2}}-\Gamma^{R}_{L-\frac{3}{2}})$. We will
look at the possible terms we will be coming across in the process
of taking the limits of the Dirac operators. In the case of spin
$\frac{3}{2}$, we see the following term:
\beq\label{49}{L(\frac{(\vec{\Sigma}.\vec{L}^L)^{3}-(\vec{\Sigma}.\vec{L}^R)^{3}}{L^{3}})}\eeq
There is also a constant factor multiplying this. However this is
not important for us right now as we are formulating rules for
taking continuum limits of such terms.

 Let us see how to take this continuum limit. For this consider
\beq{\frac{(\vec{\Sigma}.\vec{L}^L)^{3}}{L^{2}}=\frac{1}{L^2}(\vec{\Sigma}.\vec{\mathcal{L}}+\vec{\Sigma}.\vec{L}^R)^{3}}\eeq
\beq{=\frac{1}{L^{2}}[(\vec{\Sigma}.\vec{\mathcal{L}})^{3}+(\vec{\Sigma}.\vec{L}^R)^{2}(\vec{\Sigma}.\vec{\mathcal{L}})+\{\vec{\Sigma}.\vec{\mathcal{L}},\vec{\Sigma}.\vec{L}^R\}\vec{\Sigma}.\vec{\mathcal{L}}+(\vec{\Sigma}.\vec{\mathcal{L}})^{2}\vec{\Sigma}.\vec{L}^R+(\vec{\Sigma}.\vec{L}^R)^{3}+\{\vec{\Sigma}.\vec{\mathcal{L}},\vec{\Sigma}.\vec{L}^R\}\vec{\Sigma}.\vec{L}^R].}\eeq
Here we have written $\vec{L}^L=\vec{\mathcal{L}}+\vec{L}^R$ where
$\vec{\mathcal{L}}$ is the first order differential operator in the
continuum.
 In the previous equation we note that the
$(\vec{\Sigma}.\vec{L}^R)^3$ term cancels the
$-(\vec{\Sigma}.\vec{L}^R)^3$ in equation Eq.(\ref{49}). When
$L\rightarrow\infty$, the order 1 terms in $\vec{L}^R$ go away. The
$(\vec{\Sigma}.\vec{\mathcal{L}})^{3}$ also goes away in the
continuum as we take the limit. So we are left with the following
terms that have a non-zero limit
\beq{\frac{(\vec{\Sigma}.\vec{L}^L)^{3}}{L^2}=\frac{1}{L^2}[\{\vec{\Sigma}.\vec{\mathcal{L}},(\vec{\Sigma}.\vec{L}^R)^2\}+(\vec{\Sigma}.\vec{L}^R)(\vec{\Sigma}.\vec{\mathcal{L}})(\vec{\Sigma}.\vec{L}^R)].}\eeq
This is the following self-adjoint operator in the continuum:
\beq{\{\vec{\Sigma}.\vec{\mathcal{L}},(\vec{\Sigma}.\hat{x})^2\}+(\vec{\Sigma}.\hat{x})(\vec{\Sigma}.\vec{\mathcal{L}})(\vec{\Sigma}.\hat{x}).}\eeq

  The other terms we find in the expression for
the fuzzy Dirac operator for the spin $\frac{3}{2}$ case involve
powers of $\vec{L}^L$ and $\vec{L}^R$ less than $3$ and their
continuum limits were already found while we evaluated the
corresponding continuum limits in the spin $1$ and the spin
$\frac{1}{2}$ case.

At this point we make a crucial observation that the limits we are
taking are all independent of the algebra of the spin matrices
$\vec{\Sigma}$. This is the reason why we need not bother about the
order $1$ and $2$ terms in the spin $\frac{3}{2}$ case, though the
spin matrices $\vec{\Sigma}$ are different from those in the spin
$1$ case.

We are interested in finding the limits of expressions of the form
Eq.(\ref{49}), which are similar in the case of all spins, but with
higher powers of $\vec{L}^L$ and $\vec{L}^R$.

Consider first
\beq{\frac{(\vec{\Sigma}.\vec{L}^L)^4-(\vec{\Sigma}.\vec{L}^R)^4}{L^3}=\frac{1}{L^3}\left((\vec{\Sigma}.(\vec{\mathcal{L}}+\vec{L}^R))^4-(\vec{\Sigma}.\vec{L}^R)^4\right)}\eeq
\beq{=\frac{1}{L^3}\left([(\vec{\Sigma}.\vec{\mathcal{L}})^2+(\vec{\Sigma}.\vec{L}^R)^2+\{\vec{\Sigma}.\vec{L}^R,\vec{\Sigma}.\vec{\mathcal{L}}\}][(\vec{\Sigma}.\vec{\mathcal{L}})^2+(\vec{\Sigma}.\vec{L}^R)^2+\{\vec{\Sigma}.\vec{L}^R,\vec{\Sigma}.\vec{\mathcal{L}}\}]-(\vec{\Sigma}.\vec{L}^R)^4\right)}\eeq
In the above expression, only the order $3$ terms in $L^{R}$ have a
non-zero continuum limit. The $(\vec{\Sigma}.\vec{L}^R)^4$ term
cancels just as it did in expression Eq.(\ref{49}). The terms with
non-zero limit are
 \beq\frac{1}{L^3}[{\{\vec{\Sigma}.\vec{L}^R,\vec{\Sigma}.\vec{\mathcal{L}}\}(\vec{\Sigma}.\vec{L}^R)^2+(\vec{\Sigma}.\vec{L}^R)^2\{\vec{\Sigma}.\vec{L}^R,\vec{\Sigma}.\vec{\mathcal{L}}\}]}\eeq
As $L\rightarrow\infty$ this term goes to the following non zero,
self-adjoint expression
\beq{\{\vec{\Sigma}.\vec{\mathcal{L}},(\vec{\Sigma}.\hat{x})^3\}+\{\vec{\Sigma}.\vec{\mathcal{L}},\vec{\Sigma}.\hat{x}(\vec{\Sigma}.\vec{\mathcal{L}})\vec{\Sigma}\hat{x}\}}\eeq

 Looking at this pattern and using the fact that we are just
applying the binomial expansion in this computation, we can write a
general rule for computing the continuum limit for order $n$ terms.
For this we consider
\beq{\frac{1}{L^{n-1}}[(\vec{\Sigma}.\vec{L}^L)^n-(\vec{\Sigma}.\vec{L}^R)^n]}\eeq
 Again we write $\vec{L}^L=\vec{\mathcal{L}}+\vec{L}^R$ and expand
$(\vec{\Sigma}.\vec{L}^L)^n$ using the binomial expansion. As in
previous cases the $(\vec{\Sigma}.\vec{L}^R)^n$ term gets canceled
and we need to pick only the order $n-1$ terms in $\vec{L}^R$ as
these are the only terms having a non-zero continuum limit. Since
the continuum operator has to be self-adjoint and the terms
occurring in the expansion are all those occurring in a binomial
expansion, it is easy to see that the terms having a non-zero limit
can be given as the following sum:
\beq{\frac{1}{L^{n-1}}\Bigg(\sum_{k=0}^{n-1}(\vec{\Sigma}.\vec{L}^R)^{n-1-k}(\vec{\Sigma}.\vec{\mathcal{L}})(\vec{\Sigma}.\vec{L}^R)^k\Bigg)}\eeq
It is clear from this expression that we only have terms of order
$n-1$ in $\vec{L}^R$ here and we immediately see the continuum limit
of this expression as
\beq{\sum_{k=0}^{n-1}(\vec{\Sigma}.\hat{x})^{n-1-k}(\vec{\Sigma}.\vec{\mathcal{L}})(\vec{\Sigma}.\hat{x})^k.}\eeq

 Thus when considering the expression for the Dirac operator for any spin $j$, the highest order
term in $\vec{\Sigma}.\vec{L}^L$ has a power $n=2j$ and other terms
decrease from $2j$ to $1$. We have just seen how to take the
continuum limit of each of these terms with our general rules. We
also encounter polynomials in $L$ in these expressions whose limits
are easy to take. Apart from grouping terms of similar order as in
Eq.(\ref{49}), we will also encounter $\vec{\Sigma}.\vec{L}^{L,R}$
of various orders which cannot be grouped as in Eq.(\ref{49}). The
process of taking limits for such terms is straightforward and we
will not elaborate them here.

\subsection*{Verifying for spin $\frac{1}{2}$}
 We see that the highest order term is $n=1$ and so the Dirac
operator in the continuum consists of just
$\vec{\Sigma}.\vec{\mathcal{L}}$. Then we have a polynomial in
$\vec{L}^{L,R}$ in the next order whose limit combined with the
limit of the first order term gives
$\vec{\Sigma}.\vec{\mathcal{L}}+1$ as before, where
$\vec{\Sigma}=\frac{\vec{\sigma}}{2}$.

\subsection*{Verifying for spin 1}
 For spin $1$, we have $n=2$ as the highest order term and this gives the term
$\{\vec{\Sigma}.\hat{x},\vec{\Sigma}.\vec{\mathcal{L}}\}$ according
to our general rule. We should then look at the $n=1$ term which
gives a term proportional to $\vec{\Sigma}.\vec{\mathcal{L}}$. By
taking the continuum limit of
$L(\frac{\Gamma^L_{L-1}-\Gamma^R_{L+1}}{2})$ according to the rules
from the previous section, we get
\beq\label{unitaryequiv}{D'=\frac{\{\vec{\Sigma}.\vec{\mathcal{L}},\vec{\Sigma}.\hat{x}\}}{2}-(\frac{\vec{\Sigma}.\vec{\mathcal{L}}}{2}-\frac{(\vec{\Sigma}.\hat{x})^2}{2}+1)+\vec{\Sigma}.\hat{x}.}\eeq
This operator is $-\frac{1}{2}$ times the Dirac operator got in
Eq.(\ref{s1d-1}). The constant factor of $-\frac{1}{2}$ can be
absorbed in the scale factor multiplying the fuzzy Dirac operator.
In a similar way the other continuum Dirac operators can be got by
taking the limits of the correct fuzzy versions.

This verifies the rules we formulated for the know cases of spin $1$
and spin $\frac{1}{2}$.

\subsection*{Showing Unitary Equivalences}
 We speculate that $D'$ is unitarily equivalent to the Dirac
operator $D$ we got in Eq.(\ref{S1D}). We are not however able to
exactly prove it. The basis of our speculation is the unitary
equivalence of the Dirac operators of ~\cite{KP} and ~\cite{WW}
proved in ~\cite{ApbTrg}. Following that approach we consider the
following unitary transformation by the unitary operator generated
by the chirality operator $\gamma$:
\beq{D'=\left\{\exp{i\theta\gamma}\right\}D\left\{\exp{-i\theta\gamma}\right\}}\eeq
It follows from $\gamma^2=1$ and $\{D,\gamma\}=0$, that the previous
equation can be written as
\beq{D'=\left\{\exp{2i\theta\gamma}\right\}D}\eeq which is
\beq\label{63}{D'=\cos2\theta D+i\sin2\theta\gamma D.}\eeq
Substituting for $\gamma$ from Eq.(\ref{Ch1}) or Eq.(\ref{Ch-1}) we
calculate the second term in Eq.(\ref{63}) to check the equivalence.
We get \beq{\gamma D =
\left[(\vec{\Sigma}.\hat{x})^2-\vec{\Sigma}.\hat{x}\right]\vec{\Sigma}.\vec{\mathcal{L}}+(\vec{\Sigma}.\hat{x})^2-\vec{\Sigma}.\hat{x}-D.}\eeq
Most of the terms in Eq.(\ref{unitaryequiv}) are seen in the above
expression except
$(\vec{\Sigma}.\hat{x})^2\vec{\Sigma}.\vec{\mathcal{L}}$. This term
can be simplified using Eq.(\ref{19}) to get:
\beq{(\vec{\Sigma}.\hat{x})^2\vec{\Sigma}.\vec{\mathcal{L}}=\frac{1}{2}\vec{\Sigma}.\vec{\mathcal{L}}+i\epsilon_{ijm}Q_{km}\hat{x}_k\hat{x}_i\mathcal{L}_j.}\eeq

 Unitary equivalence will consist in picking $\theta$ so that
Eq.(\ref{63}) becomes Eq.(\ref{unitaryequiv}). Unfortunately we see
terms of the form
$(\vec{\Sigma}.\hat{x})^2\vec{\Sigma}.\vec{\mathcal{L}}$ in
Eq.(\ref{63}) which are not present in Eq.(\ref{unitaryequiv}).
Perhaps we must make an additional unitary transformation with a
unitary operator commuting with $\gamma$. We do not know what such
an operator can be.

We can construct more Dirac operators in the continuum starting from
the one given in Eq.(\ref{GD}). We do this by first observing that
all choices of $\gamma$ in the continuum depend only on
$\vec{\Sigma}.\hat{x}$. Hence if $P$ is a function of a variable
$\eta$ with a convergent power series expansion in $\eta$, then
\beq{ D^P =\{P(\vec{\Sigma}.\hat{x})(\Sigma_i
-\gamma\Sigma_i\gamma)+(\Sigma_i
-\gamma\Sigma_i\gamma)P(\vec{\Sigma}.\hat{x})^{\dagger}\}(\mathcal{L}_i+\Sigma_i)}\eeq
is also self-adjoint, anti-commutes with $\gamma$ and is hence also
a Dirac operator.

It is not clear if different choices of $P$ lead to unitarily
equivalent Dirac operators (after an overall scaling) or not. A
definitive answer to such questions can be obtained by calculating
the spectrum of these operators. Since we are not able to do so
analytically, we are now doing so numerically~\cite{S1DO2}.

\section{Summary of rules for finding the fuzzy Dirac operator}
\subsection*{Half-Integral Spins}
 In this case, we have an even number of projectors and hence an even
number of chiralities in the continuum. We can easily find all the
chiralities in the continuum as they are just got from constructing
projectors to various spaces labeled by the eigenvalues of
$\vec{\Sigma}.\hat{x}$.

Next we list the projectors in the fuzzy case and construct the
corresponding GW systems for each of them. So we have tables similar
to the ones in Eq.(\ref{P1})-Eq.(\ref{P-1}). Then we consider the
construction of the correct combination of the generators of the
various GW systems, which go to the chiralities found in the
continuum previously, as we take the continuum limit.

The claim is: The chiralities got from the projectors to the spaces
labeled by $j$ and $-j$ in the continuum are got by taking the
continuum limits of
\beq{\frac{\Gamma^L_{L+j}+\Gamma^R_{L-j}}{2}}\eeq and
\beq{\frac{\Gamma^L_{L-j}+\Gamma^R_{L+j}}{2}}\eeq respectively.

We now prove this claim:

Consider spin $j$ coupling to the orbital part $l$. Then if we
project to the $l+j-k$ space, it is easy to see that
\beq{Spectrum~~of~~\vec{\Sigma}.\vec{L}^L\in
lj+\frac{k}{2}[k-1-2l-2j]}\eeq where $k=0,1,...,2j$.  We use this
spectrum to construct the projectors to the above spaces.

It then follows from definition that
\beq{\frac{\Gamma^L_{l+j}+\Gamma^R_{l-j}}{2}=P^L_{l+j}+P^R_{l-j}-1}\eeq
where $P^{L,R}$ denotes the left or right projector to the
corresponding space, indicated in the suffix. Taking the continuum
limit, we get
\beq{\lim_{l\rightarrow\infty}P^L_{l+j}+P^R_{l-j}-1=\prod_{k=1}^{2j}\frac{(\vec{\Sigma}.\hat{x}-j+k)}{k}+\prod_{k=0}^{2j-1}\frac{(-\vec{\Sigma}.\hat{x}-j+k)}{(-2j+k)}-1.}\eeq
Pulling out the minus signs in the second expression we get
\beq{\lim_{l\rightarrow\infty}P^L_{l+j}+P^R_{l-j}-1=\frac{\prod_{k=1}^{2j}(\vec{\Sigma}.\hat{x}-j+k)}{(2j)!}+(-1)^{4j}\frac{\prod_{k=0}^{2j-1}(\vec{\Sigma}.\hat{x}+j-k)}{(2j)!}-1.}\eeq
Since $4j$ is even for both integral and half-integral $j$,
observing that
$\prod_{k=1}^{2j}\vec{\Sigma}.\hat{x}-(j-k)=\prod_{k=0}^{2j-1}\vec{\Sigma}.\hat{x}+(j-k)$,
we get
\beq{\lim_{l\rightarrow\infty}\frac{\Gamma^L_{l+j}+\Gamma^R_{l-j}}{2}=2\frac{\prod_{k=1}^{2j}(\vec{\Sigma}.\hat{x}-j+k)}{(2j)!}-1.}\eeq
This is exactly the expression for the chirality operator got in the
continuum from the projector to the space where
$\vec{\Sigma}.\hat{x}=j$.

 Now since, $L\left(\frac{\Gamma^L_{L+j}-\Gamma^R_{L-j}}{2}\right)$ and
$\frac{\Gamma^L_{L+j}+\Gamma^R_{L-j}}{2}$ anticommute in the fuzzy
case, they will continue to do so as we take the continuum limit. So
we can be sure that
\beq{L\left(\frac{\Gamma^L_{L+j}-\Gamma^R_{L-j}}{2}\right)}\eeq
gives us the fuzzy Dirac operator corresponding to this chirality.

 We can follow the same procedure to get the remaining fuzzy Dirac
and chirality operators, exhausting all possibilities.

\subsection*{Integral Spins}
In this case, we have an odd number of projectors and hence an odd
number of chiralities in the continuum. We then proceed as we did
for the case of half-integral spins and we note that all the
arguments go through, except when it comes to the Dirac operator
corresponding to the chirality obtained from the projector to the
space where $\vec{\Sigma}.\hat{x}=0$. In this case, we construct the
fuzzy analogues from the generators of the GW system obtained from
the left and right projectors to the $L+0$ space alone. We cannot
mix the generators of the GW system got from this projector with the
generators obtained from the projectors to other spaces as we get
diverging continuum limits. We omit the simple details for showing
this result.

\section{Index theory for the spin $j$ Dirac operator}

 The index of the Dirac operator can be computed by counting the
number of zero modes. These zero modes are eigenstates of the Dirac
operator spanning a subspace left invariant by the chirality
operator. Thus if chirality is diagonalised in this subspace of zero
modes and the dimensions of the zero mode subspaces with $\gamma=\pm
1$ are $n_{L,R}$, the index of the Dirac operator is $n_L-n_R$.
There will be a minimum of $n_L-n_R$ linearly independent zero modes
of the Dirac operator with $\gamma=1$($\gamma=-1$), if $n_L\geq n_R$
$(n_L\leq n_R)$, respectively.

We can compute the index as follows ~\cite{Fuzzy,ApbSv}. Consider
the instanton sectors of $S^2$, which correspond to $U(1)$ bundles
thereon. On $S_F^2$, projective modules substitute for sections of
bundles.

We build the projective modules on $S_F^2$ by introducing a spin $T$
representation of $SU(2)$ whose carrier space is
$\mathbb{C}^{2T+1}$. We then consider,
$Mat(2L+1)\otimes\mathbb{C}^{2T+1}$, on which $SU(2)$ acts with
generators $\vec{L}^L+\vec{T}$. Then we consider,
$Mat(2L+1)\otimes\mathbb{C}^{2T+1}\otimes\mathbb{C}^{2j+1}$, the
space where the fuzzy spin $j$ Dirac operator with instanton
coupling acts. The desired projective modules are then constructed
by considering $P^{L\pm
T}Mat(2L+1)\otimes\mathbb{C}^{2T+1}\otimes\mathbb{C}^{2j+1}$, where
$P^{L\pm T}$ is the projector to the space where $\vec{L}^L+\vec{T}$
couple to $L+T$ and $L-T$ respectively. The different projectors
obtaining by varying $T$ as well correspond to different Chern
numbers which classify the projective modules in the continuum and
in the fuzzy case. Using these projectors we can construct their
corresponding GW systems and hence the fuzzy Dirac operators with
instanton coupling. We do not explicitly show the construction of
the projective modules for a general spin $j$ here. For details
regarding spin $\frac{1}{2}$, see~\cite{Fuzzy}.

Next we find the unpaired eigenstates obtained by combining the four
angular momenta, namely $\vec{L}^L,\vec{T},-\vec{L}^R,\vec{\Sigma}$,
to get the total angular momentum $\vec{J}$. Unpaired eigenstates
are those whose eigenvalues are got by combining the four angular
momenta in a unique way. These are eigenstates of the total angular
momentum $\vec{J}$. These are also eigenstates of the Dirac operator
as $\vec{J}$ commutes with the Dirac operator. The method of
counting the number of unpaired eigenstates is illustrated in
Table(\ref{sometable}), where we have considered the case where
$\vec{\Sigma}=\frac{1}{2}$. We note that the states with total
angular momentum $T-\frac{1}{2}$ and $2L+T+\frac{1}{2}$ are the
unpaired ones as they occur just once in Table(\ref{sometable}). The
latter is the top mode and we can discard it as it does not agree
with the values obtained in the continuum ~\cite{Index}, (See Page
95, Chapter 8 of ~\cite{Fuzzy}). We are then left with the space
whose value of total angular momentum is $T-\frac{1}{2}$ and the
dimension of this space is $2T$. This is the number of zero modes of
the Dirac operator and hence its index. This space is left invariant
by the chirality operator.

\begin{table}
\begin{center}
\begin{tabular}{|c|c|c|}
\hline
$\vec{L}^L+\vec{T}-\vec{L}^R$ & $\vec{\Sigma}$ & $\vec{J}$ \\
\hline \hline
$0+T \rightarrow$ & $-\frac{1}{2}\rightarrow$ & $T-\frac{1}{2}$  \\
$0+T \rightarrow$ & $\frac{1}{2}\rightarrow$ & $T+\frac{1}{2}$  \\
$1+T \rightarrow$ & $-\frac{1}{2} \rightarrow$ & $T+\frac{1}{2}$ \\
$1+T \rightarrow$ & $\frac{1}{2} \rightarrow$  & $T+\frac{3}{2}$ \\
$2+T \rightarrow$ & $-\frac{1}{2} \rightarrow$ & $T+\frac{3}{2}$ \\
\vdots & \vdots & \vdots \\
$2L+T-1 \rightarrow$ & $\frac{1}{2}\rightarrow$ & $2L+T-\frac{1}{2}$  \\
$2L+T \rightarrow$ & $-\frac{1}{2} \rightarrow$ & $2L+T-\frac{1}{2}$ \\
$2L+T \rightarrow$ & $+\frac{1}{2} \rightarrow$ & $2L+T+\frac{1}{2}$\\
 \hline
\end{tabular}
\end{center}
\caption{Method to find unpaired eigenstates} \label{sometable}
\end{table}

This procedure can be carried out for any spin $j$. When we do this,
we find that the only unpaired eigenstate, discarding the top mode,
is the one with the eigenvalue $T-j$. This is also the minimum value
of the total angular momentum. This gives us $2(T-j)+1$ as the
number of zero modes and this is the index of these Dirac operators.

We can verify for the familiar~\cite{Fuzzy} spin $\frac{1}{2}$ case
that this gives $2T$. For the case of spin $1$, this gives $2T-1$.

\section{Conclusions}
 We have seen that we can construct Dirac operators for any spin on
the fuzzy $2$-sphere. We made use of the properties of the
projectors to various spaces to achieve this construction and by
formulating rules to take their continuum limits we found these
operators on the commutative $2$-sphere as well. A general
construction of the Dirac and chirality operators on the continuum
$2$-sphere was shown.

Formulating the gauge sectors of these operators in the fuzzy case
\cite{Fuzzy} and taking their continuum limits, we can also get
equations with interactions on $S^2$ and $S_F^2$.

 We can construct the Dirac and chirality operators on
$\mathbb{R}^2$. We did not show this construction here as it is
quite straight forward and can be done using our general methods for
constructing them. Moreover we did not obtain any new result by
considering them.

We are examining the spectrum of these Dirac operators on $S^2$ and
$S_F^2$. We could not find them analytically and so we are trying to
do it numerically~\cite{S1DO2}. We are also studying quantum field
theories associated with these operators by functional integral
techniques.

\section{Acknowledgements}
 We thank Prof.T.R.Govindarajan for the helpful discussions and the
support he gave one of us (PP) at IMSc,Chennai. We also thank
Prof.Sachin Vaidya for useful discussions and references. We also
thank Anosh Joseph, Earnest Akofor and M.Martone for helpful
discussions.

 PP thanks Prof.Sachin Vaidya for his kind hospitality in IISc, Bengaluru where
this work was started. APB thanks Alberto Ibort and the Universidad
Carlos III de Madrid for their kind hospitality and support.

 The work was supported in part by DOE under the grant number DE-FG02-85ER40231.
The work of APB was also supported by the Department of Science and
Technology, India.

\bibliographystyle{apsrmp}

\begin{thebibliography}{99}

\bibitem{Fuzzy} A.P. Balachandran, S. Kurkcuoglu, S. Vaidya, {\it Lectures on fuzzy and fuzzy SUSY
physics}, World Scientific Publishing(2007).

\bibitem{AC} A. Connes, {\it Noncommutative Geometry}, Academic Press, London, 1994

\bibitem{KP} H. Grosse, C. Klimcík, P. Prešnajder, {\it Topologically nontrivial field configurations in noncommutative
geometry}, Commun.Math.Phys.507(1996) and hep-th/9510083.

\bibitem{Do} B.P. Dolan, I. Huet, S. Murray, D.O. Connor, {\it Noncommutative vector bundles over fuzzy $\mathbb{C}P^N$ and their covariant
derivatives}, JHEP(2007) and  hep-th/0611209, 2006.

\bibitem{GW} P. H. Ginsparg and K. G. Wilson, {\it A Remnant of Chiral Symmetry on the Lattice},
Phys. Rev. D25, 2649 (1982).

\bibitem{ApbYd} S. Baez, A.P. Balachandran, S. Vaidya, B. Ydri, {\it Monopoles and solitons in fuzzy
physics}, Commun.Math.Phys.787-798(2000) and hep-th/9811169v6.

\bibitem{ApG} G. Alexanian, A.P. Balachandran, G. Immirzi, B. Ydri, {\it Fuzzy $\mathbb{C}P^N$}, Jour.Geom.Phys., Volume 42, Issues 1-2, May 2002, Pages 28-53.

\bibitem{ApGi} A.P. Balachandran, Giorgio Immirzi, {\it The Fuzzy Ginsparg-Wilson Algebra: A Solution of
the Fermion Doubling Problem}, Phys.Rev. D68 (2003) 065023 and
hep-th/0301242v2.

\bibitem{ApbTrg} A. P. Balachandran, T. R. Govindarajan, B.
Ydri, {\it The Fermion Doubling Problem and Noncommutative
Geometry}, Mod.Phys.Lett. A15 (2000) 1279 and hep-th/9911087v2; {\it
Fermion doubling problem and noncommutative geometry II},
hep-th/0006216v1.

\bibitem{WW} U. Carow-Watamura and S. Watamura, {\it Chirality and Dirac Operator on Noncommutative
Sphere}, Commun. Math. Phys. 183 (1997) 365 and hep-th/9605003; {\it
Noncommutative Geometry and Gauge Theory on Fuzzy Sphere}, Commun.
Math. Phys. 212 (2000) 395 and hep-th/9801195.

\bibitem{Jaya} Camillus Jayawardena, {\it Schwinger Model on $S^2$},
Helvetica Physica Acta, Vol 61(1988) 636-711.

\bibitem{Va} D.A. Varshalovich, A.N. Moskalev, V.K. Khersonskii,
{\it Quantum Theory of Angular Momentum}, World Scientific(1987).

\bibitem{S1DO2} A.P. Balachandran, Santan Digal, Pramod Padmanabhan,
{\it In Preparation}.

\bibitem{ApbSv} A.P. Balachandran, S. Vaidya, {\it Instantons and Chiral Anomaly in Fuzzy Physics}, hep-th/9910129.

\bibitem{Index} A. Bassetto and L. Griguolo, {\it Chiral anomalies for vortex potentials in two dimensions and a decompactification limit}, Journ. of Math. Phys. 32 (1991) 3195.




\end{thebibliography}

\end{document}